\documentclass[aps,physrev,preprint,groupedaddress,amsmath,amssymb]{revtex4-2}

\usepackage{graphicx}
\usepackage{dcolumn}
\usepackage{bm}
\usepackage{tabularx}
\usepackage[flushleft]{threeparttable}
\usepackage{multirow}
\usepackage{xspace,xfrac}
\usepackage{subfig}
\makeatletter
\usepackage{hyperref}
\newcommand\footnoteref[1]{\protected@xdef\@thefnmark{\ref{#1}}\@footnotemark}
\makeatother
\usepackage{ifthen} 
\newboolean{articletitles}
\setboolean{articletitles}{true} 
\usepackage{fix-cm}
\usepackage{cleveref}
\usepackage{anyfontsize}

\newcommand{\GeV}{\ensuremath{~\text{GeV}}\xspace}
\newcommand{\TeV}{\ensuremath{~\text{TeV}}\xspace}

\newcommand{\ETmiss}{\ensuremath{E_{\mathrm{T}}^{\text{miss}}}\xspace}
\newcommand{\dPhi}{\ensuremath{\Delta\phi}\xspace}
\newcommand{\ZPrime}{\ensuremath{Z^{\prime}}\xspace}
\newcommand{\mZPrime}{\ensuremath{m_{\ZPrime}}\xspace}
\newcommand{\Pt}{\ensuremath{p_{\mathrm{T}}}\xspace}
\newcommand{\pT}{\ensuremath{p_{\mathrm{T}}}\xspace}
\newcommand{\PtISR}{\ensuremath{\Pt^{\text{ISR}}}\xspace}

\newcommand{\drap}{\ensuremath{y*}\xspace}
\newcommand{\dR}{\ensuremath{\Delta R}\xspace}
\newcommand{\rinv}{\ensuremath{r_{\text{inv}}}\xspace}

\newcommand{\rinvML}{\ensuremath{r_{\mathrm{inv}}^{\mathrm{ML}}}}
\newcommand{\rinvAna}{\ensuremath{r_{\mathrm{inv}}^{\mathrm{analytic}}}}

\newcommand{\maos}{\ensuremath{m_{\text{MAOS}}}\xspace}

\newcommand*{\MGMCatNLOV}[1]{\textsc{MadGraph5}\_aMC@NLO~#1\xspace}
\newcommand*{\PYTHIAV}[1]{\textsc{Pythia}~#1\xspace}
\newcommand*{\DELPHES}[1]{\textsc{Delphes}~#1\xspace}

\begin{document}

\title{\textbf{Reconstructing the Invisible Fraction of Semi-visible Jets in ISR-Boosted Events via Neural Network Regression}}%

\author{Yin Li}
\affiliation{School of Science, Shenzhen Campus of Sun Yat-sen University, 66 Gongchang Road, Shenzhen, Guangdong 518107, PRC}

\author{Bingxuan Liu}%
\email{Contact author: liubx28@mail.sysu.edu.cn}
\affiliation{%
School of Science, Shenzhen Campus of Sun Yat-sen University, 66 Gongchang Road, Shenzhen, Guangdong 518107, PRC
}%

\author{Jian Bin Wang}
\affiliation{School of Science, Shenzhen Campus of Sun Yat-sen University, 66 Gongchang Road, Shenzhen, Guangdong 518107, PRC}

\author{Jia Qi Xie}
\affiliation{School of Science, Shenzhen Campus of Sun Yat-sen University, 66 Gongchang Road, Shenzhen, Guangdong 518107, PRC}

\author{Kai Rong Xu}
\affiliation{School of Science, Shenzhen Campus of Sun Yat-sen University, 66 Gongchang Road, Shenzhen, Guangdong 518107, PRC}

\author{Rui Han Ye}
\affiliation{School of Science, Shenzhen Campus of Sun Yat-sen University, 66 Gongchang Road, Shenzhen, Guangdong 518107, PRC}

\author{Zi Huan Huang}
\affiliation{School of Science, Shenzhen Campus of Sun Yat-sen University, 66 Gongchang Road, Shenzhen, Guangdong 518107, PRC}

\date{\today}
             
\begin{abstract}
Semi-visible jets (SVJs) provide a characteristic collider signature of strongly interacting dark sectors, in which the key model parameter \rinv controls the fraction of dark hadrons decaying to dark matter candidates. In this work, a regression model is developed to reconstruct \rinv in SVJ events produced in association with an energetic photon. The model uses information from high-level physics objects only, and the training procedure is optimized to ensure applicability. The performance is found to be robust against varying signal parameters and \rinv can be reconstructed at a much higher precision, compared to previously developed analytical method. It offers a new approach to conduct SVJ searches that can potentially unify both $s$-channel and $t$-channel productions, enhancing the sensitivities.
\end{abstract}

\maketitle
\clearpage

\section{Introduction}
\label{sec:intro} 

The Standard Model of particle physics has achieved great success, but it does not contain a particle that can serve as a dark matter (DM) candidate. It is therefore necessary to introduce new physics, in particular dark matter models beyond the Standard Model (SM), given the strong hints of its existence~\cite{Rubin:1980Rotation,Persic:1995URC,Clowe:2006DarkMatterProof,DES:2017WeakLensing,Planck:2020Results}. One of the main candidates for DM is the weakly interacting massive particle (WIMP)~\cite{WIMP}. Although searches of this type have covered a wide range of parameter space, they are focused on the scenario where the DM candidate is produced in isolation~\cite{monojet-atlas, monojet-cms, ATLAS:2024fdw,CMS:2024zqs}. This picture can be qualitatively modified if the dark sector has its own confining dynamics. In the dark QCD framework, analogous to quantum chromodynamics (QCD) in the SM, the hard interaction produces dark quarks ($\chi_i$), which subsequently shower and hadronize under a new confining force $SU(N)$~\cite{Strassler:2006im, Han:2007ae}. The resulting dark hadrons can have different detector signatures, where some remain stable and invisible on detector scales, while others decay back to SM particles. Therefore, the missing transverse momentum is not produced only as an isolated recoil object, but can be embedded within and geometrically correlated with jet-like visible activity. This gives rise to a striking experimental signature in which a significant fraction of the energy inside the jets is invisible. Such jets are called semi-visible jets (SVJs)~\cite{Cohen:2015SVJ,Cohen:2017DarkShowers,Beauchesne:2018myj, Beauchesne:2017yhh}.

Previous studies of SVJ used the transverse mass, $m_T$, constructed from the dijet system and missing transverse energy, \ETmiss, as the core discriminating variable to gain sensitivity to the $s$-channel $Z' \to \chi\bar\chi$ process~\cite{Cohen:2015SVJ}. The experimental signature is determined by a few effective parameters that dominate the observable behaviour, among which \rinv controls the ratio of invisible components inside the jets~\cite{Cohen:2017DarkShowers}. The community has also attempted to consolidate the parameter choices to ensure smooth cross-talks~\cite{Albouy:2022DarkShowers}. Beyond the canonical vector-mediator setup, it has also been studied for more general mediator structures carrying both SM and hidden-sector charges~\cite{Beauchesne:2017yhh}. Both the ATLAS and CMS experiments have probed the $s$-channel SVJ productions recently~\cite{CMS:2021ResonantDarkMatter,ATLAS:2025kuz}, following the technique proposed in ref.~\cite{Cohen:2015SVJ}. Although the $t$-channel production lacks a clear resonant mass peak, it is possible to take advantage of the event topology since the alignment of jets and \ETmiss is unique, as reported by a dedicated ATLAS search in this channel~\cite{ATLAS:2023SVJSearch}. Efforts have been made to explore the distinguishable features within SVJ using substructure observables~\cite{Kar:2020SVJSubstructure,Park:2017rfb,Cohen:2020afv,Cohen:2023mya} or constituent-level information~\cite{Faucett:2022LearningSVJ,Luigi:SVJModels,Bernreuther:2020vhm}. Recent studies have further advanced machine-learning approaches for SVJ toward more model-agnostic self-supervised methods and symmetry-aware graph-network architectures~\cite{Luigi:SVJModels,Canelli:2021aps,Bhardwaj:2024djv}. 

Along with the identification of SVJ, the reconstruction of key physical quantities has also been developed in parallel. Existing studies investigated the application of machine-learning methods with an information-compression structure to learn the mediator mass from event dynamics~\cite{Pedro:2023OptimalMassSVJ}. At the same time, initial state radiation (ISR) in SVJ events is found to be beneficial. High-$p_T$ ISR objects help with event triggering and background suppression, provide an additional boost to the SVJ system, and change the geometric relationship between \ETmiss\ and SVJ~\cite{ATLAS:2024qqm}. Assuming a simple collinear condition that the invisible energy in each SVJ aligns with the jet axis allows us to reconstruct \rinv\ analytically~\cite{Liu:2024SVJplusX}. These attempts are of great interest as they do not only provide discriminating variables, but also enable direct extraction of the model parameters using real data, in case of a discovery. In particular, \rinv\ is the key parameter that governs the phenomenologies in both $s$-channel and $t$-channel productions. If one can reconstruct \rinv\ more precisely, it is feasible to consider both channels simultaneously. The positive results from ref.~\cite{Liu:2024SVJplusX} indicate that it can be achieved using event-level variables already, which is naturally more robust against varying model parameters. 

\begin{figure}[t]
    \centering
    \includegraphics[width=\linewidth]{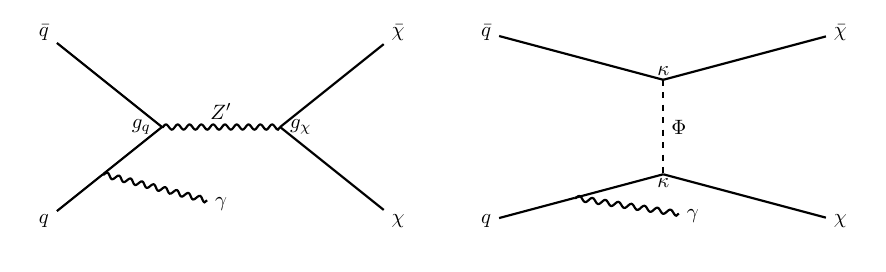}
    \caption{Representative Feynman diagrams of the signal processes with an ISR photon considered in this work. Left: the $s$-channel process, \(q\bar q \to Z' + \gamma_{\mathrm{ISR}}\), followed by \(Z' \to \chi\bar{\chi}\). Right: the $t$-channel process, \(q\bar q \to \chi\bar{\chi} + \gamma_{\mathrm{ISR}}\), mediated by a scalar \(\Phi\).}
    \label{fig:sch_tch_feynman}
\end{figure}

This work tries to establish a workflow to improve the \rinv reconstruction performance, in production channels with an energetic ISR photon, as shown in Fig.~\ref{fig:sch_tch_feynman}. We demonstrate that by setting up a neural network with sophisticated data pre-processing, it is possible to achieve a much better precision. The performance is robust against various model parameters and effective to both $s$- and $t$-channel production channels. The findings in this article reveals the potential of event-topology in SVJ searches, and offer practical strategies that can be applied in future experimental searches.

The article is structured as follows: Section~\ref{sec:dataset} describes the datasets used for this study; Section~\ref{sec:obs} introduces the event slection criteria and various key variables; Section~\ref{sec:model} lays out the training workflow and algorithm structure, followed by a study to evaluate its impact in Section~\ref{sec:impact}, and finally Section~\ref{sec:conclusion} concludes this work with some discussion points.   
\section{Datasets}
\label{sec:dataset}

All events are generated using Monte Carlo (MC) simulations. In this study, the center-of-mass energy is assumed to be $\sqrt{s}$ =13\TeV, corresponding to the data-taking conditions of the LHC Run~2. The hard scattering processes are generated with \MGMCatNLOV{3.5.1}~\cite{madgraph}. Parton showering and hadronization are performed using \PYTHIAV{8.306}~\cite{pythia}. The dark sector dynamics are implemented through the built-in \texttt{Hidden Valley} module in \PYTHIAV. The output is subsequently passed to \DELPHES{3.5.3}~\cite{delphes} for fast detector simulation, using the CMS detector geometry. Jets are clustered using the anti-$k_t$ algorithm~\cite{Cacciari:2008AntiKt,Cacciari:2011FastJet} with a jet radius of $R=0.4$.

\subsection{Signal processes}
\label{sec:signal}

Three sets of signal samples are generated to suit different tasks, including the algorithm training, robustness testing and generality testing. Detailed information is given below.

\subsubsection{Algorithm training samples}
\label{sec:training_sample}

The training samples consider a \ZPrime boson produced via an $s$-channel process, decaying into a pair of dark quarks, $\chi$, i.e.,\ $Z' \to \chi\bar\chi$. The details of the model can be found in ref.~\cite{CMS:2021ResonantDarkMatter}. Since this work studies the case with an energetic ISR object, we require at least one ISR photon at the generation level. Processes with an additional parton are also included. The signal grid spans across \ZPrime mass and \rinv, summarized in Table~\ref{tab:signal_parameters}, together with other important model parameters. Each signal point contains 100k events.  The benchmark grid is chosen to cover the main parameter dependence relevant for the photon-associated SVJ topology. The \ZPrime mass range defines the main high-mass training region, while the \rinv scan from $0.1$ to $0.9$ covers the semi-visible regime between the mostly visible and mostly invisible limits. The other dark-sector parameters define the nominal Hidden Valley configuration used for training, with variations studied separately in Sec.~\ref{sec:impact}. The Hidden Valley sector is configured with $N_C^{\rm dark}=2$ and $N_F^{\rm dark}=2$.  The degenerate choice $m_{\pi_D}=m_{\rho_D}=20$\GeV\ is treated as a simplified generator-level benchmark rather than a full dark-hadron spectroscopy model.

\begin{table}[htbp]
\centering
\caption{Main parameters used in the benchmark signal event generation. The
corresponding benchmark choices are discussed in the text.}
\label{tab:signal_parameters}
\begin{tabular}{lc}
\hline
Parameter & Value \\
\hline
Mediator mass $m_{Z'}$ & $1500$--$3500$\GeV\\
Center-of-mass energy $\sqrt{s}$ & $13$\TeV\\
Invisible fraction \rinv & $0.1$--$0.9$ (step $0.1$) \\
Dark colour number $N_C^{\rm dark}$ & $2$ \\
Dark flavour number $N_F^{\rm dark}$ & $2$ \\
Dark quark mass $m_{\chi}$ & $10$\GeV\\
Dark pion mass $m_{\pi_D}$ & $20$\GeV\\
Dark rho mass $m_{\rho_D}$ & $20$\GeV\\
Dark confinement scale $\Lambda_D$ & $35.1539$\GeV\\
Vector meson probability \texttt{probVector} & $0.75$ \\
Events per signal point & $100\,000$ \\
\hline
\end{tabular}
\end{table}

\subsubsection{Robustness testing samples}
\label{sec:robust_sample}

To validate the robustness of the trained model, additional datasets are generated under different dark sector model assumptions. For $s$-channel SVJ events, the baseline parameter configuration is chosen to be \[ \texttt{probVector} = 0.75,\quad \Lambda = 35.1539\GeV,\quad m_{\pi_D} = m_{\rho_D} = 20\GeV.  \]
Single-parameter variations are then performed for the following quantities:

\begin{itemize} 

\item the dark vector meson probability (\texttt{probVector}), with values
of $0.5$ and $0.6$; 
\item the dark confinement scale ($\Lambda$), with values of $10$ and $50$\GeV; 
\item the dark pion and dark rho masses ($m_{\pi_D}$ and $m_{\rho_D}$), set to $10$ and $40$\GeV.
\item the dark pion masses ($m_{\pi_D}$), set to $100$ and $200$\GeV.
\item the dark flavour number ($N_F^{\mathrm{dark}}$), with values of $3$ and $4$.
\end{itemize}

In addition, to test the dependence on the mediator mass beyond the main training grid, extra samples with $\mZPrime = 500$ and $1000\GeV$ are generated and evaluated separately. To ensure that variations in the machine learning model performance can be attributed to changes in a specific dark sector parameter rather than to correlated multi-parameter effects, only one parameter is varied at a time in each trial, while all other parameters are fixed to their baseline values.

\subsubsection{Generality testing samples}
\label{sec:generality_sample}

In addition to the resonant $s$-channel production mode, SVJs can also be produced through a non-resonant $t$-channel production. As shown in Fig.~\ref{fig:sch_tch_feynman}, the connection between SM and the dark sector is established by a scalar mediator $\Phi$, which couples a quark in SM to a dark quark through a Yukawa-type interaction. The partonic process in proton--proton collisions can produce dark quarks via $t$-channel mediator exchange, which then undergo dark showering and hadronization to form SVJs. A set of $t$-channel samples is prepared to examine whether the algorithm trained on the $s$-channel remains effective. The same event generation, detector simulation, and reconstruction procedure as for the $s$-channel signal is adopted. For two different $\Phi$ mass points at 1.5, 2.5 and 4.0\TeV, samples are generated at three \rinv values, \rinv = 0.1, 0.5, and 0.9.

\subsection{Background}
\label{sec:background}

The background process used for the algorithm training is $\gamma$+jets.  Since its cross-section falls rapidly with increasing jet transverse momentum, a slicing strategy based on the minimum transverse momentum of the leading jet is adopted to cover the high-momentum region relevant for the signal without losing statistical precision.  Specifically, subsamples are generated with the minimum leading jet transverse momentum $p_{T,\mathrm{jet}_1}^{\mathrm{min}}$ starting from $0$ to $2000$\GeV, in steps of $500$\GeV. Each slice contains $40{,}000$ unweighted events. The background samples are reconstructed and preselected using the same procedures as those applied to the signal samples, ensuring that all datasets share a consistent phase-space definition and kinematic treatment before being fed into the training process.

\section{Selections and Observables}
\label{sec:obs}

This section discusses the selections applied to mimic actual analyses, and introduces the definition of the training target. A brief summary on the analytical method from ref.~\cite{Liu:2024SVJplusX} is also given. 

\subsection{Event Selections}
\label{sec:preproc}

We follow the same strategy as Ref.~\cite{Liu:2024SVJplusX} for preselection. Events are required to contain at least one photon with transverse momentum greater than $150~\GeV$. After removing the overlapping photons and jets, at least two reconstructed jets are required. The two leading jets are then selected as the SVJ candidates for the construction of the observables used later. The corresponding selection criteria are summarized in Table~\ref{tab:event_selection}. In particular, to retain a broader event sample for the subsequent observable study and regression task, only a loose requirement of  $|y^{*}|<5$ is imposed in this work.



\begin{table}[htbp]
\centering
\caption{Object definitions and event selection criteria used in this analysis.}
\label{tab:event_selection}
\begin{tabular}{lc}
\hline
Selection requirement & Value \\ 
\hline
Photon multiplicity & $N_{\gamma} \geq 1$ \\ 
Jet multiplicity & $N_{\mathrm{jet}} \geq 2$ \\

ISR transverse momentum & $\PtISR > 150\GeV$ \\ 
Jet transverse momentum& $\Pt^{j_1} , \Pt^{j_2} > 25\GeV$\\

Photon–jet separation & $\dR(\gamma,\mathrm{jet}) > 0.4$ \\ 
Jet azimuthal separation & $|\dPhi(\mathrm{jet}_1,\mathrm{jet}_2)| > 0.8$ \\ 
Rapidity separation & $|\drap| < 5$ \\

\hline
\end{tabular}
\end{table}

\subsection{\rinv\ Training Target}
\label{sec:rinv_label}

Following a commonly adopted definition in LHC SVJ studies, \rinv is defined as
\begin{equation} \rinv = \frac{N_{\mathrm{invisible}}}{N_{\mathrm{visible}} +
N_{\mathrm{invisible}}} \label{eq1}
\end{equation}~\cite{Albouy:2022DarkShowers}, where $N_{\mathrm{invisible}}$
($N_{\mathrm{visible}}$) refers to the number of dark hadrons decaying to detector invisible (visible) particles.  For different jets with the same particle-level  \rinv parameter, the relative fractions of visible and invisible components can still fluctuate following a binomial law. Therefore, we establish a method to obtain the event-by-event \rinv value. 

It is calculated as \begin{equation} \rinv = \frac{N_{\mathrm{D
M}}}{2(N_{\pi_{D}} + N_{\rho_{D}})} \label{eq2}
\end{equation} 
and is used as the target variable in the machine learning training. In the actual implementation, as shown in the formula, the numerator is taken to be the number of final-state dark matter particles, identified by selecting particles with $\mathrm{PID}=51$ or $53$ and $\mathrm{status}=1$. The denominator is constructed from dark mesons, including both dark pion and dark rho mesons. No distinction is made between diagonal and off-diagonal dark mesons in this definition; all hadronized dark pion and dark rho states recorded by Pythia are included.  The former are identified by $|\mathrm{PID}|=4900111$ or $\pm 4900211$, while the latter are identified by $|\mathrm{PID}|=4900113$ or $4900213$, and in both cases the particles are required to have $\mathrm{status}=83$ or $84$. The factor two in the denominator comes from the fact that a dark hadron either decays to a pair of DM particles, or a pair of SM quarks in our setup. Fig.~\ref{fig:analytic_reco} illustrates the particle-level  \rinv distributions of a few signal points.

\subsection{Analytical \rinv Reconstruction}
\label{sec:rinv_reco}

In a realistic experimental environment, generator-level information is not directly accessible, and \rinv must be indirectly estimated using reconstruction-level observables. When a high transverse-momentum ISR object (either a jet or a photon) recoils against the SVJ system, the event-topology becomes clearer. Previous studies have attempted to analytically reconstruct the invisible fraction of the jets in this scenario~\cite{Liu:2024SVJplusX}.

The analytic method assumes that \ETmiss comes from the two invisible components of the SVJs, with each aligned with the corresponding jet axis. Consequently, the transverse momentum conservation of the event can be written as \begin{equation} \begin{aligned} p_x^{\mathrm{miss}} &= \alpha_1\, p_{x,1} +
\alpha_2\, p_{x,2}, \\ p_y^{\mathrm{miss}} &= \alpha_1\, p_{y,1} + \alpha_2\,
p_{y,2}, \end{aligned} \label{eq3} \end{equation} where $p_{x,i}$ and $p_{y,i}$ denote the transverse momentum components of the $i$-th jet, and $\alpha_i$ represents the ratio between the invisible and visible components within that jet. Solving this linear system yields $\alpha_1$ and $\alpha_2$, from which
\rinv for a given event can be defined as \begin{equation}
r_{\mathrm{inv}}^{\mathrm{analytic}} = \frac{1}{2} \left(
\frac{\alpha_1}{1+\alpha_1} + \frac{\alpha_2}{1+\alpha_2} \right).\label{eq4}
\end{equation} This analytically reconstructed value, \rinvAna, has a clear physical interpretation, a simple functional form, and directly exploits the global transverse momentum balance of the event. However, its performance depends strongly on the validity of the underlying assumptions. The geometric interpretation of this decomposition follows Ref.~\cite{Liu:2024SVJplusX}. In this approximation, the two leading jets are treated as the two SVJ candidates, and the missing transverse momentum is assumed to be dominated by the invisible components of these two SVJs. Each invisible component is further assumed to be approximately aligned with the corresponding jet direction. Additional radiation, imperfect SVJ assignment, detector effects, or other sources of missing transverse momentum can therefore spoil this simplified topology and degrade the analytic reconstruction.

\begin{figure}[t]
    \centering
    \includegraphics[width=0.85\linewidth]{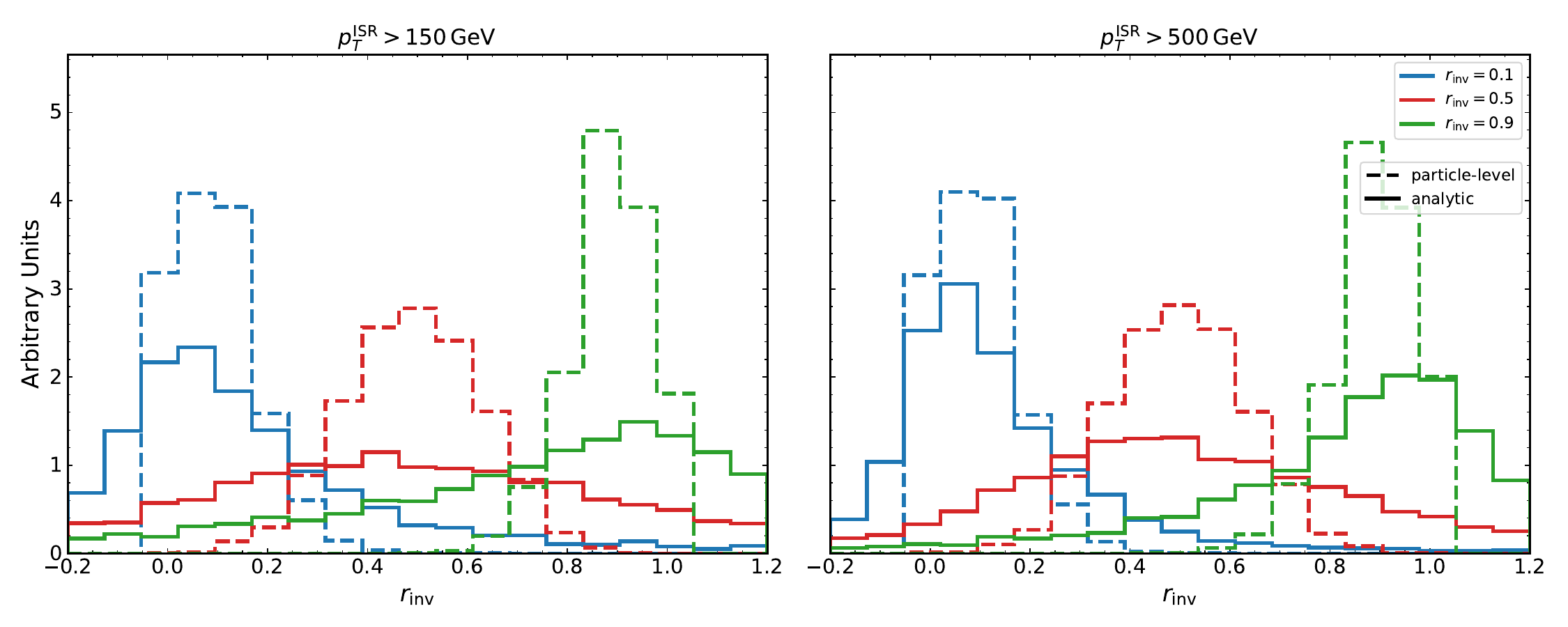}
    \caption{Distributions of the analytically reconstructed \rinv\ for events with particle-level $\rinv=0.1$, $0.5$, and $0.9$, under the selections $\PtISR>150$~\GeV\ (left) and $\PtISR>500$~\GeV\ (right). The dashed histograms show the corresponding particle-level \rinv\ distributions.
}
    \label{fig:analytic_reco}
\end{figure}

At relatively low ISR transverse momentum thresholds ($\pT > 150\GeV$), the distribution of \rinvAna typically remains broad, shows only limited correlation with the generator-level \rinv, and contains a non-negligible fraction of negative values or solutions exceeding the physically allowed range, as shown in Fig.~\ref{fig:analytic_reco}. This indicates that the geometric approximations in the analytic method are not sufficiently satisfied in this regime. When the ISR transverse momentum is increased ($\pT > 500\GeV$), the analytic reconstruction improves and the \rinvAna distribution becomes narrower, reflecting the fact that a more boosted topology better satisfies the collinear condition. Nevertheless, the \rinvAna distribution is still wider than that of \rinv computed directly from particle-level information. This suggests that a multi-variate method trained for various conditions may achieve better performance. 

\section{The Regression Model}
\label{sec:model}

In this section, we discuss the algorithm structure and the training workflow, followed by a study evaluating the performance.  

\subsection{Neural network and training strategy}
\label{sec:nn}

Since we want to explore the potential of high-level physics objects, such as jets and \ETmiss, without any low-level information such as jet substructures, a simple fully connected feed-forward neural network should suit our needs. For the input features, a set of observables that can be robustly obtained at the reconstruction level is selected. These include the four-momentum information of the ISR object, the two leading jets, and \ETmiss, as listed in Table~\ref{tab:input_observables}. This feature set provides a sufficient physical basis for the model to infer \rinv. All input features are pre-processed. In the first step, a physics-motivated normalization is applied, in which energy- and momentum-related quantities are divided by \pT of the leading jet, which is therefore not included as input feature. This reduces the dependence of the model performance on the absolute mass scale of \ZPrime and improves the robustness. Standard normalization is then applied to further stabilize the optimization process.

\begin{table}[t]
\centering
\begin{threeparttable}
\caption{Input features used in the \rinv regression.}
\label{tab:input_observables}
\begin{tabular}{ll}
\hline
Category & Observables \\
\hline
ISR object 
& $\Pt^{\gamma}$, $\eta^{\gamma}$, $\phi^{\gamma}$ \\

Leading jet 
& $\eta^{j_1}$, $\phi^{j_1}$, $m^{j_1}$ \\

Sub-leading jet 
& $\Pt^{j_2}$, $\eta^{j_2}$, $\phi^{j_2}$, $m^{j_2}$ \\

Missing transverse momentum 
& \ETmiss, $\phi^{\mathrm{miss}}$ \\
\hline
\end{tabular}
\end{threeparttable}
\end{table}

To avoid biases arising from an imbalanced \rinv distribution, a sampling procedure is employed to all signal samples listed in Section~\ref{sec:training_sample}, after applying the selections in Section~\ref{sec:preproc}. The entire \rinv region starting from zero to one is divided into 20 equal-width bins. The third-lowest bin population is chosen as a common sampling threshold, and the number of events in each bin is randomly reduced to this threshold. This procedure ensures that all bins have approximately equal statistical weights in the training dataset.

The dataset is split into training, validation, and test sets with a ratio of
$0.6:0.2:0.2$. The regression model is trained separately under different ISR and sample-composition conditions, corresponding to different physical use cases. In particular, the model is trained under two ISR momentum thresholds, $\PtISR > 150$\GeV and $\PtISR > 500$\GeV, reflecting the difference between moderately and highly boosted regimes. For each ISR selection, the training samples are further divided into two categories: samples containing only signal events, and samples composed of a mixture of signal and dominant background processes. The former is used to study the \rinv regression performance under ideal conditions, while the latter more closely resembles the practical analysis scenario in which signal and background coexist. For background events, \rinv cannot be computed using Eq.~\eqref{eq2}. To retain a continuous target variable in the mixed training sample, the background \rinv is instead assigned using Eq.~\eqref{eq4}. In total, four independent regression models are constructed. They share the same fully connected regression framework, while the network depth and training hyperparameters are chosen separately for different training setups. In the implementation used in this work, the models trained with $\PtISR > 150$\GeV employ five hidden layers, whereas those trained with $\PtISR > 500$\GeV employ four hidden layers. Further technical details of the input preprocessing, network architecture, training setup, final hyperparameters, and public inference workflow are provided in Appendix.

\subsection{Performance of \rinv regression}
\label{sec:rinv_perf}

\begin{figure}[t]
    \centering
    \includegraphics[width=0.85\linewidth]{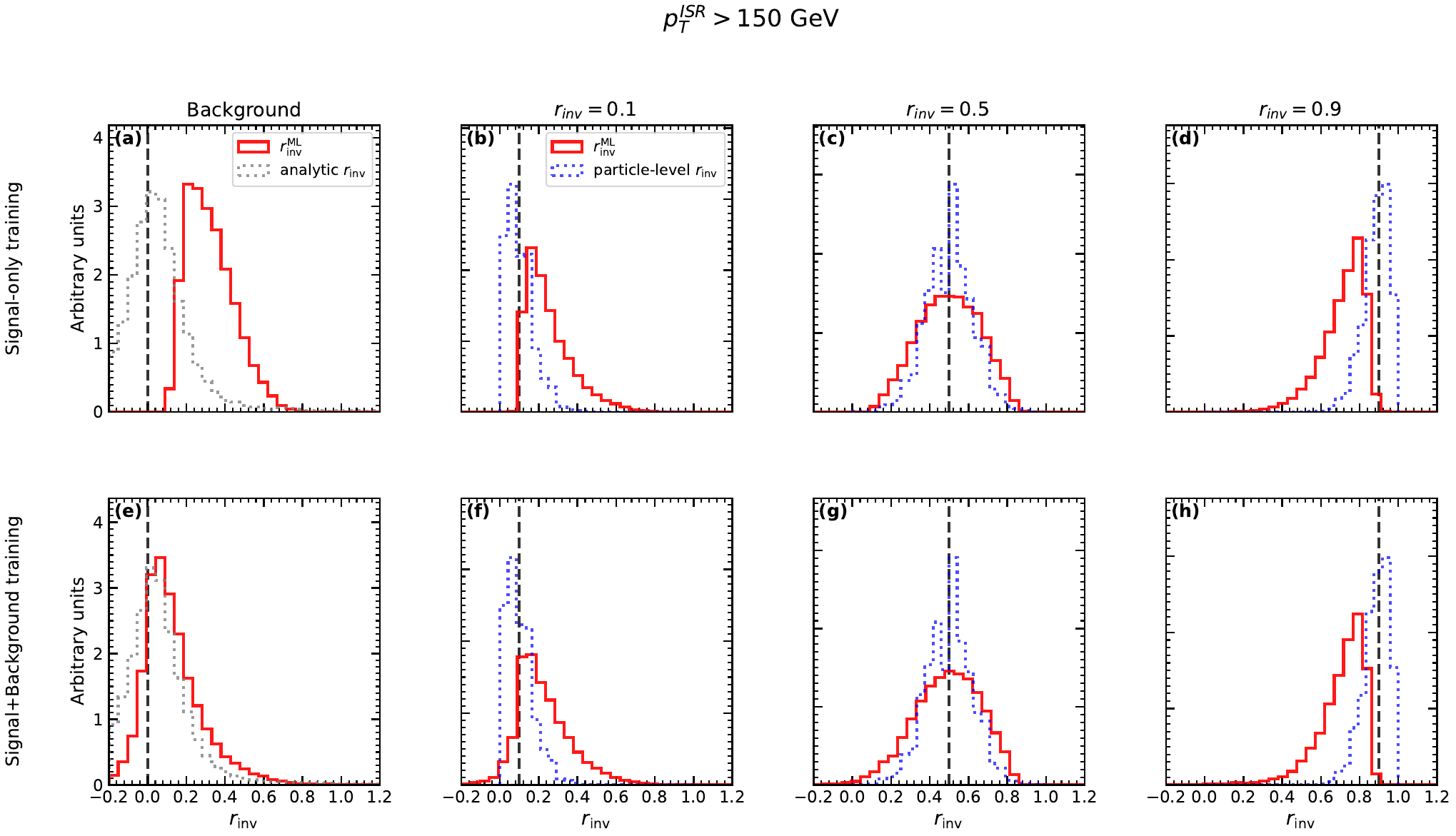}
    \caption{Distributions of \rinvML\ for the $\PtISR>150$\GeV selection. The upper row shows the models trained with signal-only samples, and the lower row shows the models trained with signal-plus-background samples. From left to right, the panels correspond to background, $r_{\mathrm{inv}}=0.1$, $0.5$, and $0.9$. The dashed histograms indicate the particle-level \rinv\ distributions.}
    \label{fig:training_pt150.pdf}
\end{figure}

\begin{figure}[t]
    \centering
    \includegraphics[width=0.85\linewidth]{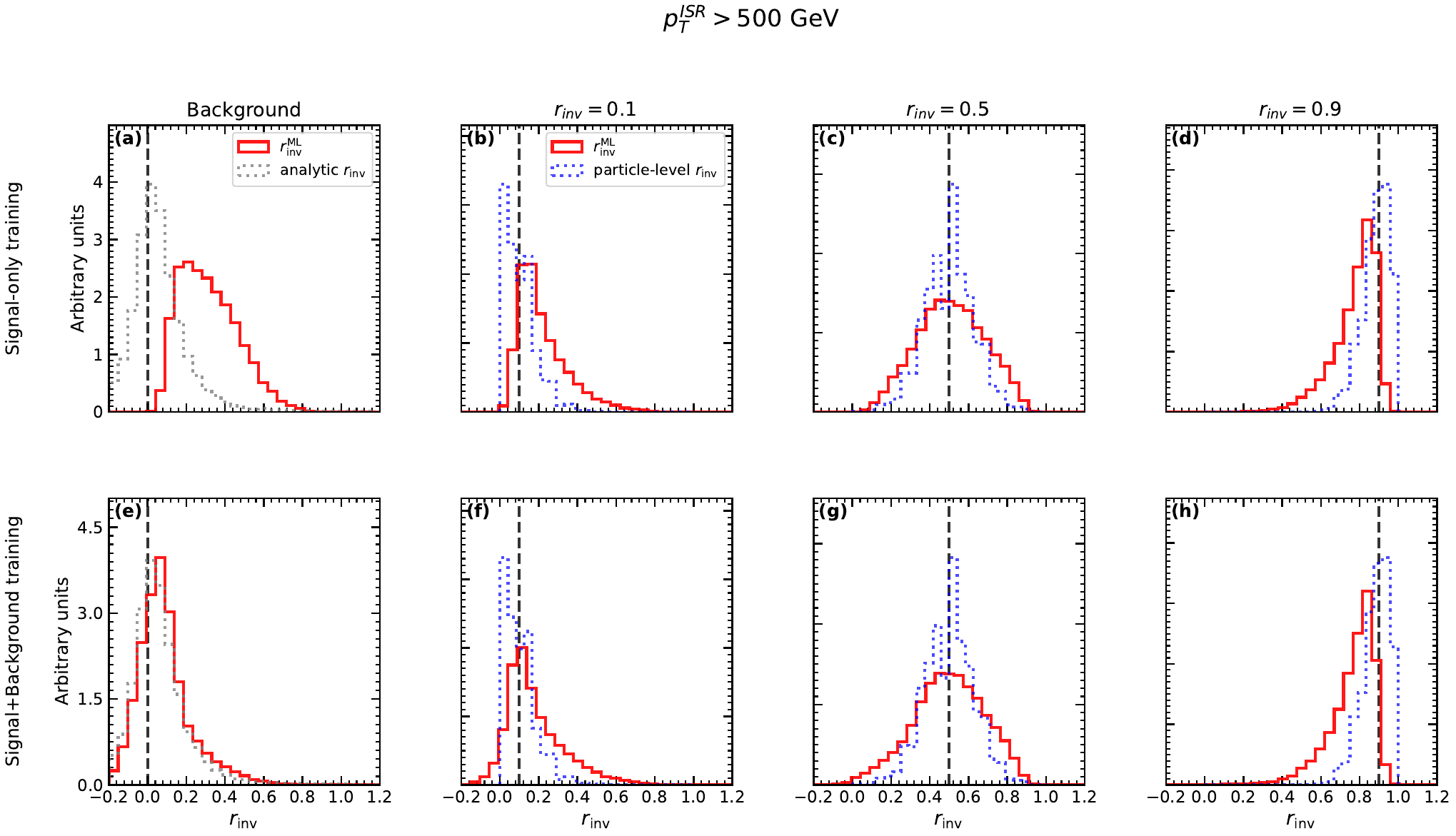}
    \caption{Distributions of \rinvML\ for the $\PtISR>500$\GeV selection. The upper row shows the models trained with signal-only samples, and the lower row shows the models trained with signal-plus-background samples. From left to right, the panels correspond to background, $r_{\mathrm{inv}}=0.1$, $0.5$, and $0.9$. The dashed histograms indicate the particle-level \rinv\ distributions.}
    \label{fig:training_pt500.pdf}
\end{figure}

The regression performance is examined under both ISR selections, $\PtISR > 150\GeV$ and $\PtISR > 500\GeV$. Fig.~\ref{fig:training_pt150.pdf} and~\ref{fig:training_pt500.pdf} show that the trained model gives an overall ordered response to the generator-level \rinv\ in all four training setups. For both ISR thresholds, the reconstructed output follows the change of the true \rinv\ consistently, indicating that the event-level observables used as inputs retain sufficient information for the regression task.

\begin{figure}[t]
    \centering
    \includegraphics[width=0.85\linewidth]{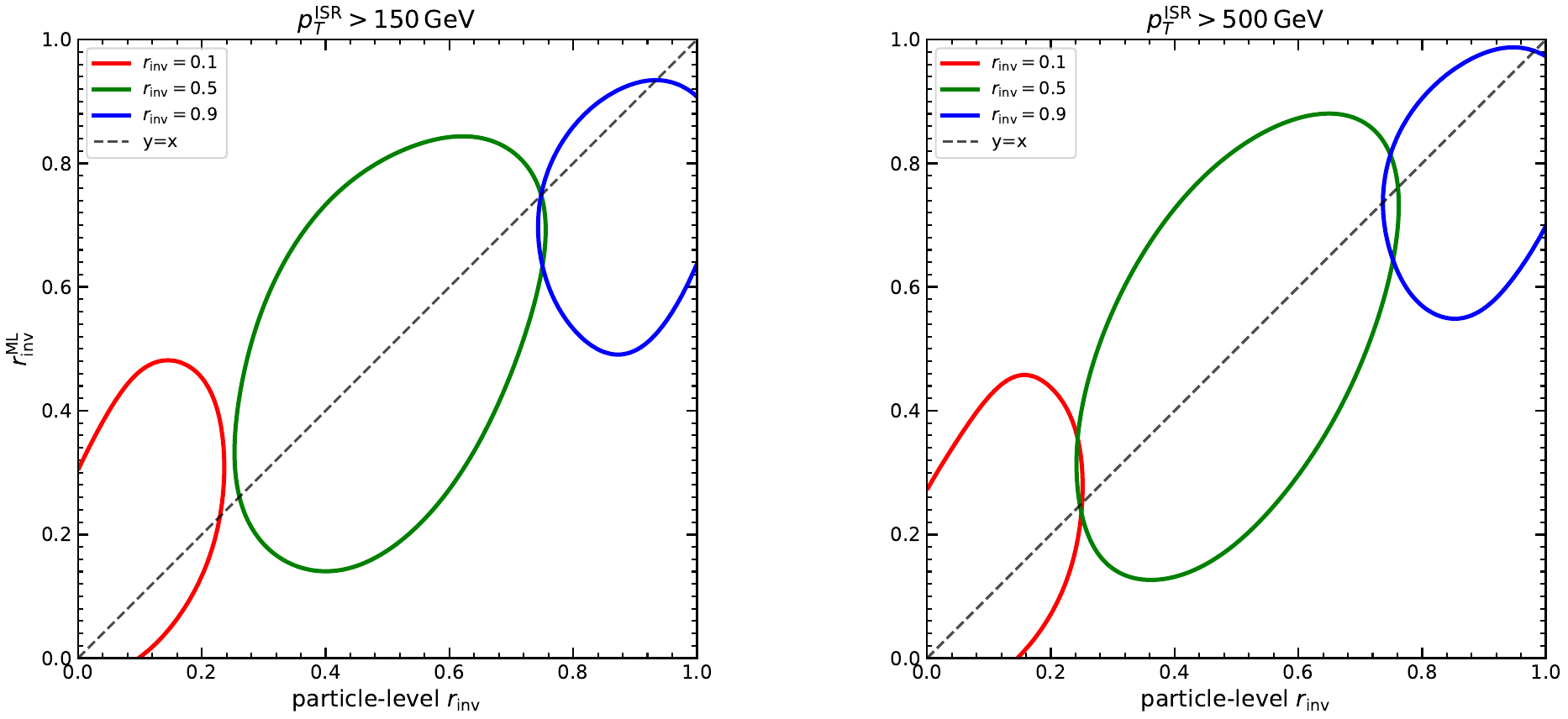}
    \caption{Two-dimensional comparison between \rinvML\ and the particle-level \rinv, shown as 80\% contour regions for $\rinv=0.1$, $0.5$, and $0.9$, under the selections $\PtISR>150$\GeV (left) and $\PtISR>500$\GeV (right). The dashed line indicates perfect reconstruction.}
    \label{fig:2drinv.pdf}
\end{figure}

\begin{figure}[t]
    \centering
    \includegraphics[width=0.85\linewidth]{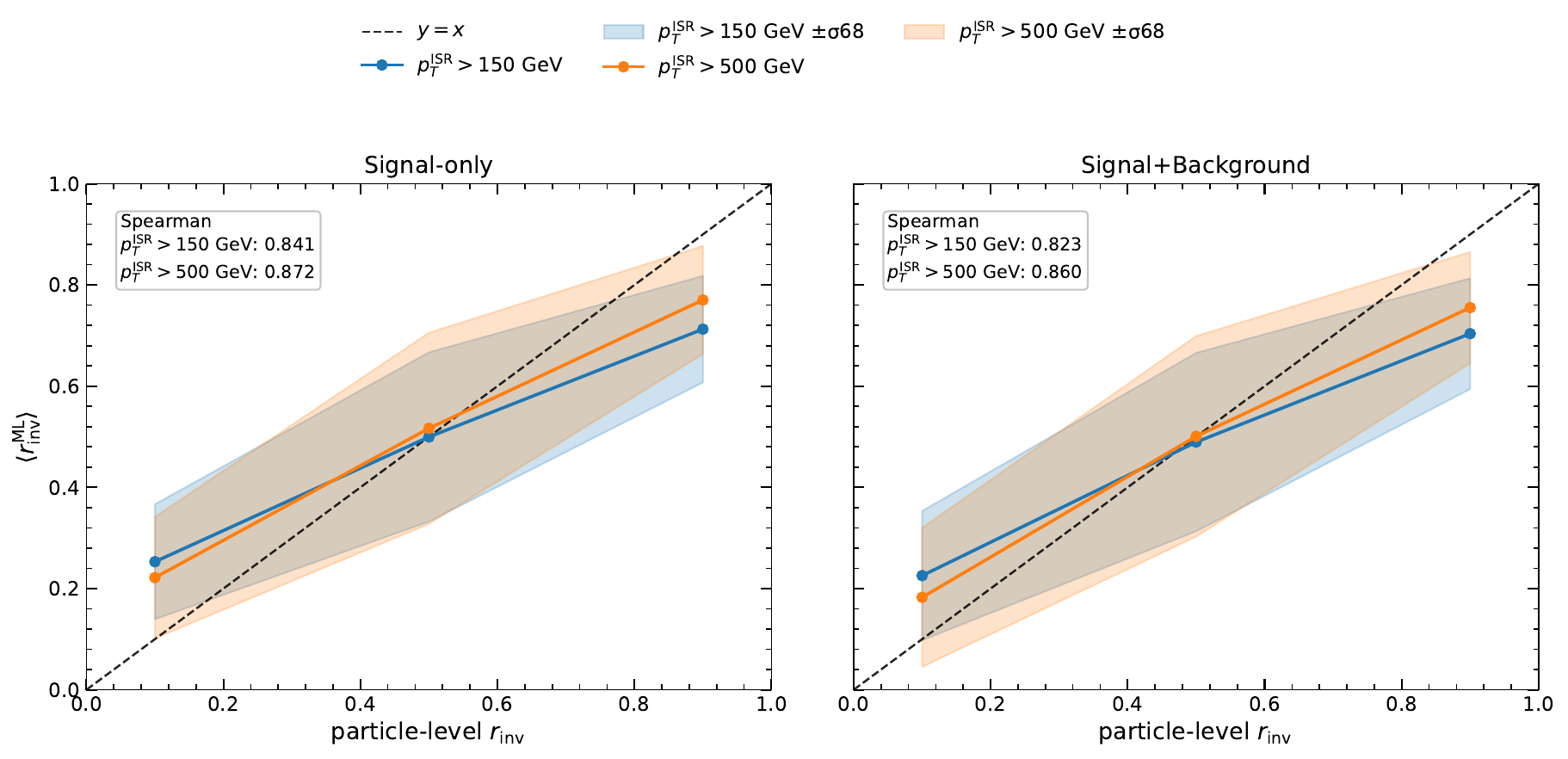}
    \caption{Mean \rinvML\ response as a function of \rinv\ for the signal-only (left) and signal-plus-background (right) training setups. Results for the two ISR selections are shown together with the corresponding $\pm\sigma$ variations. The dashed line indicates perfect reconstruction.}
    \label{fig:mlsummary}
\end{figure}

Fig.~\ref{fig:2drinv.pdf} shows that different \rinv\ hypotheses remain more clearly separated in the comparison with the particle-level label, while Fig.~\ref{fig:mlsummary} states that the inclusion of background only mildly weakens the overall behaviours and does not qualitatively change the ordered signal response. Since the two ISR selections are trained independently, the change of the ISR threshold has only a limited impact on the regression quality. 

\begin{figure}[t]
    \centering
    \includegraphics[width=0.85\linewidth]{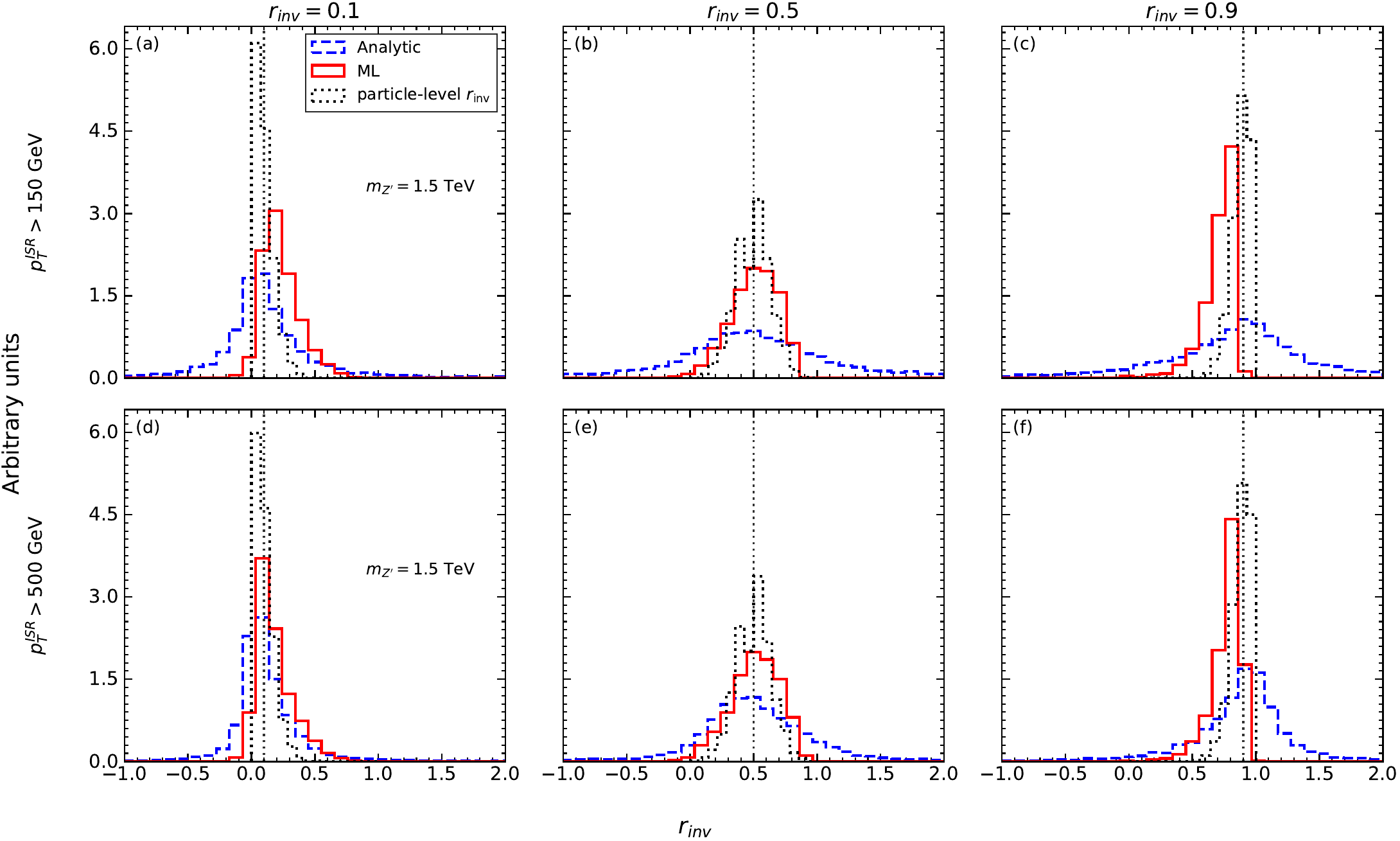}
    \caption{Comparison between the ML-regressed and analytically reconstructed \rinv\ distributions for $\mZPrime=1500$~\GeV. Panels (a)--(c) correspond to $\PtISR>150$\GeV with $\rinv=0.1$, $0.5$, and $0.9$, while panels (d)--(f) show the corresponding distributions for $\PtISR>500$~\GeV. The dashed histograms indicate the particle-level \rinv\ distributions.}
    \label{fig:ml_analytic_1500}
\end{figure}

\begin{figure}[t]
    \centering
    \includegraphics[width=0.85\linewidth]{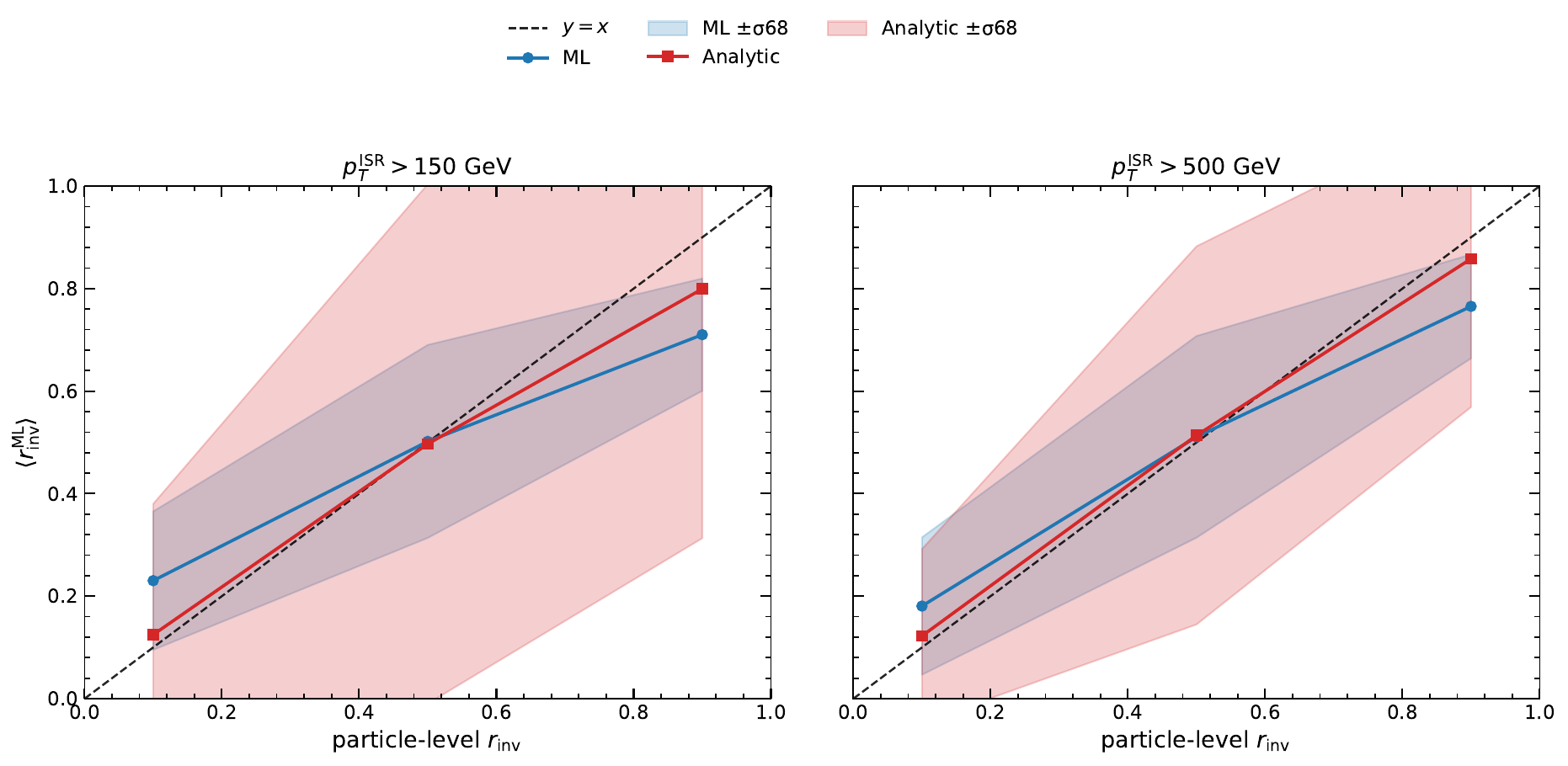}
    \caption{Mean reconstructed \rinv\ as a function of the particle-level \rinv\ for the ML and analytic methods, shown for $\PtISR>150$\GeV (left) and $\PtISR>500$\GeV (right). The shaded bands indicate the corresponding $\pm\sigma$ variations, and the dashed line corresponds to perfect reconstruction.}
    \label{fig:ml_vs_analytic_summary}
\end{figure}

As discussed in Section~\ref{sec:rinv_reco}, \rinvAna is strongly constrained by the underlying recoil and collinearity assumptions. Consequently, its reconstructed distributions are substantially broader in the low-ISR regime, as shown in Fig.~\ref{fig:ml_analytic_1500} and Fig.~\ref{fig:ml_vs_analytic_summary}, which reduces the separation power. On the contrary, the machine learning approach yields narrower overall distributions, and the algorithm can be re-trained to adapt for a new ISR condition, offering more flexibility in the analysis usage.

\clearpage

\section{Impact}
\label{sec:impact}

We evaluate the potential impact of the regression model in this section. Firstly, the signal parameter dependence is examined using varied signal samples, and then the model is tested on $t$-channel production. At last, it is  revealed that both $s$-channel and $t$-channel may benefit from \rinvML\ in practice. 

\subsection{Robustness against dark-sector parameter variations}
\label{sec:robust_para}

The physics-motivated scaling method in Section~\ref{sec:model} attempts to remove the model's dependence on the mediator masses. However, given the rich set of signal parameters listed in Table~\ref{tab:signal_parameters}, it is imperative to make sure the regression model is not over-trained towards a specific setup. Therefore, the stability of the regression results is examined under different dark-sector parameter configurations.

\begin{figure}[th]
    \centering
    \includegraphics[width=0.85\linewidth]{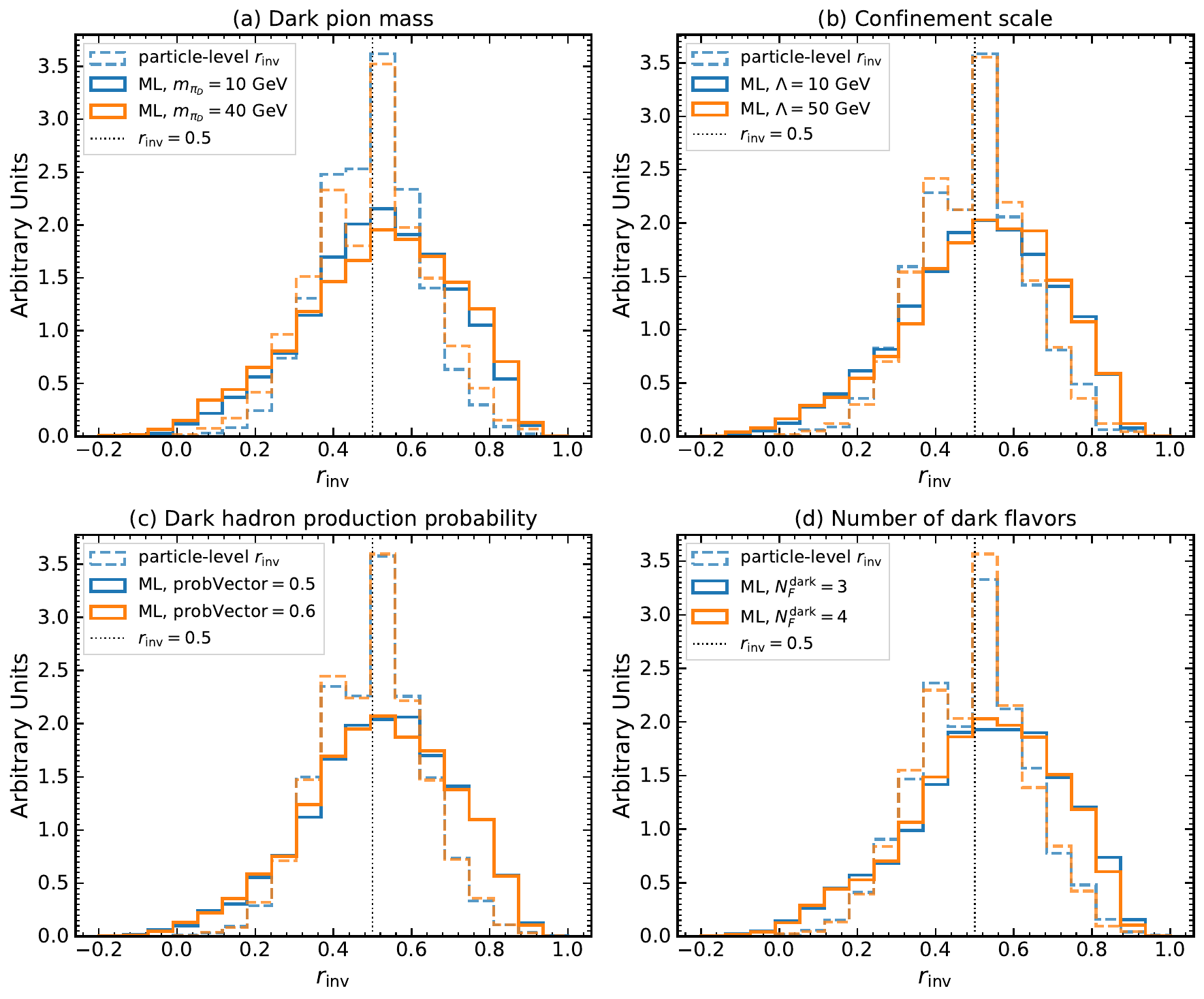}
    \caption{Regressed \rinvML\ distributions at fixed $\rinv=0.5$ for variations of the dark-sector parameters: (a) dark pion mass ($m_{\pi_{D}}$), (b) confinement scale ($\Lambda$), (c) dark hadron production probability (\texttt{probVector}), and (d) dark flavour number ($N_F^{\mathrm{dark}}$). In each panel, only the parameter indicated is varied, while the others are kept at their baseline values.}
    \label{fig:robustness2x2}
\end{figure}
\begin{figure}[th]
    \centering
    \includegraphics[width=0.45\linewidth]{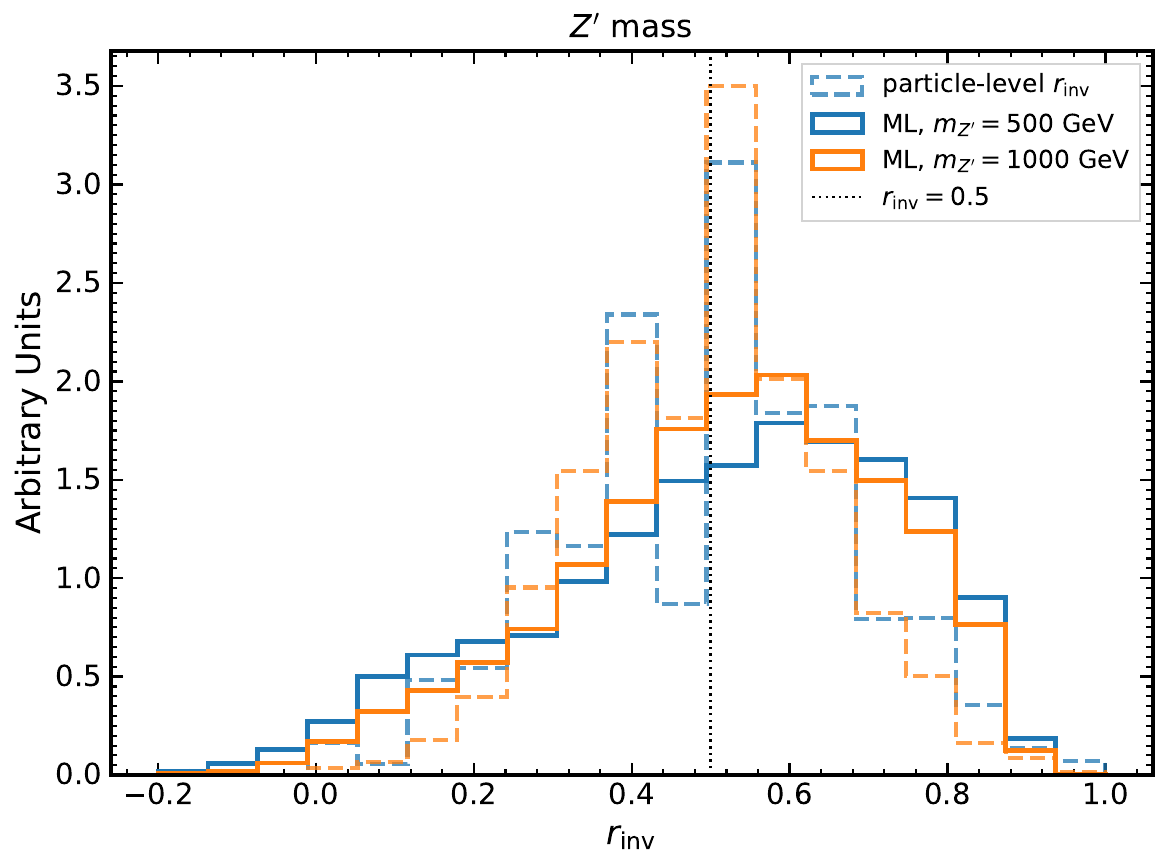}
    \caption{Regressed \rinvML\ distributions under variations of the mediator mass \mZPrime at fixed $\rinv=0.5$. The dashed histograms show the particle-level \rinv distributions, while the solid histograms show the corresponding \rinvML predictions evaluated with the baseline regression model. The vertical dotted line marks the nominal $\rinv=0.5$ value. The similar particle-level and ML distributions indicate that the regression response remains stable under the mediator-mass variation considered here.}
    \label{robustness_zmass}
\end{figure}

The regression performance is evaluated with only one dark-sector-related parameter varied at a time, while keeping the other conditions fixed. The parameters considered include the probability of dark hadron production (\texttt{probVector}), the dark confinement scale ($\Lambda$), the dark pion and dark rho masses ($m_{\pi_D}$ and $m_{\rho_D}$), the dark flavour number ($N_F^{\mathrm{dark}}$), as well as the $Z'$ mass (\mZPrime). Fig.~\ref{fig:robustness2x2} shows representative results for variations of the dark-sector parameters at fixed $\rinv=0.5$, including $m_{\pi_D}$, $\Lambda$, \texttt{probVector}, and $N_F^{\mathrm{dark}}$. Only mild changes are observed, suggesting that the model response is not strongly tied to a specific choice of these dark-sector parameters.
The mediator mass dependence is shown separately in Fig.~\ref{robustness_zmass}. This variation probes the stability of the regression response under changes in the hard-process energy scale. The predicted \rinvML distributions remain close to the corresponding particle-level distributions, indicating that the model response is also stable under the mediator-mass variation considered here.

\begin{figure}[th]
    \centering
    \includegraphics[width=0.8\textwidth]{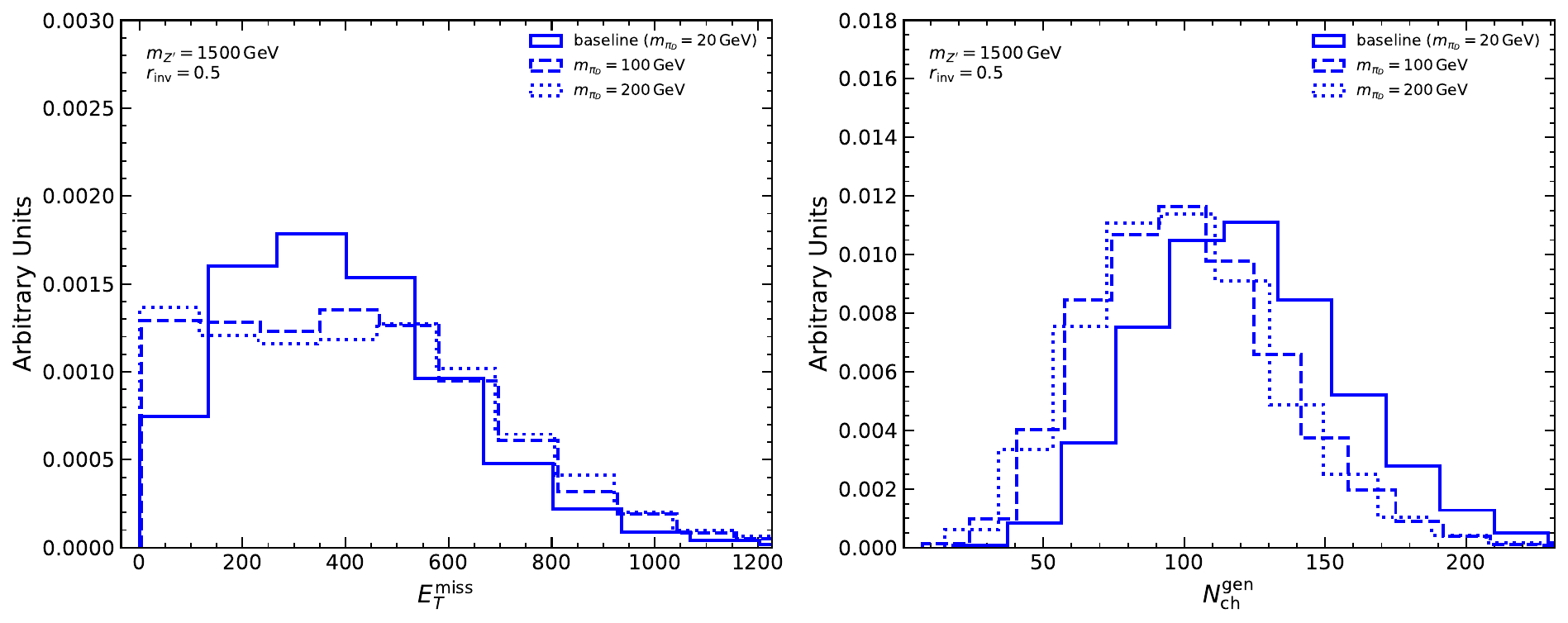}
    \caption{Left: comparison of the normalized \ETmiss\ distributions for three cases with $m_{\pi_{D}}=20\GeV$, $100\GeV$, and $200\GeV$. When the dark hadron mass is increased to relatively extreme values, the peak of the \ETmiss\ distribution is weakened and the distribution becomes flatter. Right: distribution of the generator-level charged-particle multiplicity $N_{\mathrm{ch}}^{\mathrm{gen}}$, which shifts overall to lower values as the dark meson mass increases.}
    \label{fig:met_and_nch}
\end{figure}

\begin{figure}[th]
    \centering
    \includegraphics[width=0.45\textwidth]{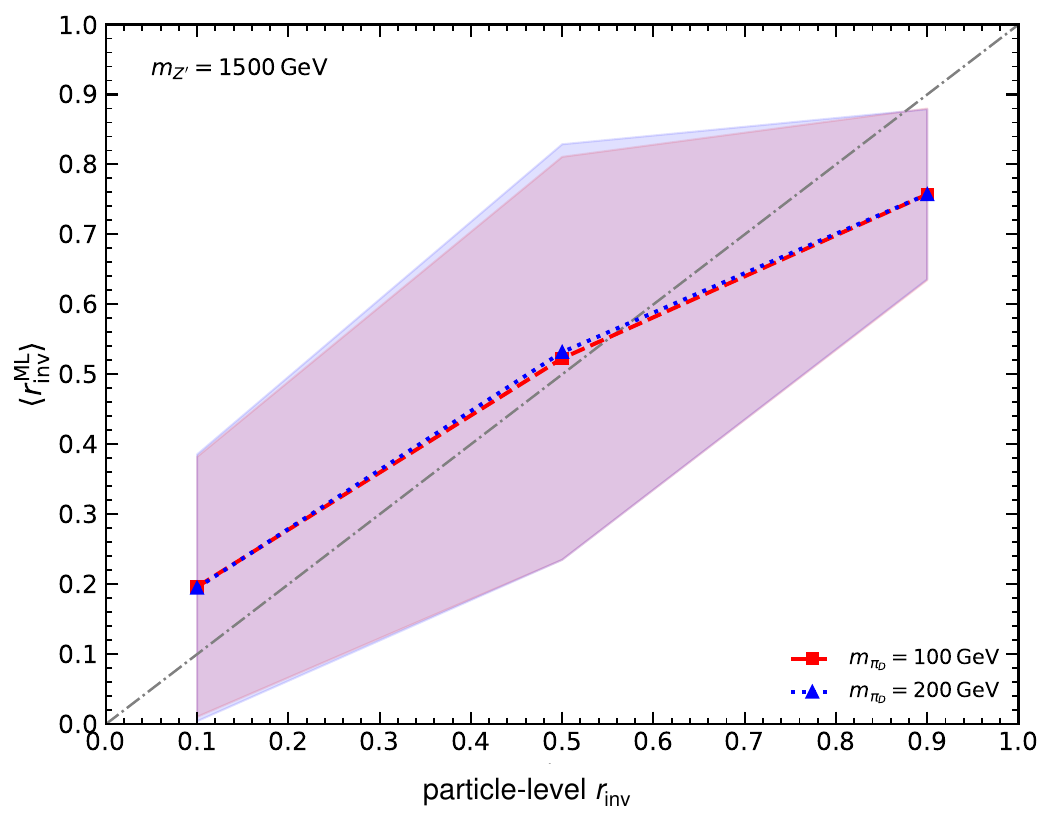}
    \caption{Mean response of \rinvML\ as a function of the true \rinv\  for the cases with increased dark meson mass. The monotonic and ordered behavior of \rinvML\ is not significantly changed when the dark meson mass is increased, indicating that the regression model remains stable under this parameter variation.}
    \label{fig:pion100and200_rinvML_mean}
\end{figure}

The results are not surprising. The boosted topology induced by a high-$p_T$ ISR object improves the resolution of the relevant observables and makes the geometric structure of the system clearer~\cite{Liu:2024SVJplusX}. In such a phase-space, those event-level quantities are not impacted by the details of dark-shower, unlike jet substructure type variables. Three $m_{\pi_D}$ points are compared here as illustrative examples, including two extreme values, 100 and 200 GeV. When $m_{\pi_D}$ is increased to such high values, the dark shower produces fewer dark hadrons, which in turn modifies shower-level observables such as shown in Fig.~\ref{fig:met_and_nch}. While the number of charged particles sees a clear shift, the \ETmiss distribution changes mildly in the bulk.  Fig.~\ref{fig:pion100and200_rinvML_mean} shows the regression performance remains broadly stable for those conditions. It is reasonable to anticipate an algorithm exploring the correlations between charged-particles may be affected to a much greater extent.


\clearpage

\subsection{Applicability to $t$-channel production }
\label{sec:robust_t}

\begin{figure}[h]
    \centering
    \includegraphics[width=0.85\linewidth]{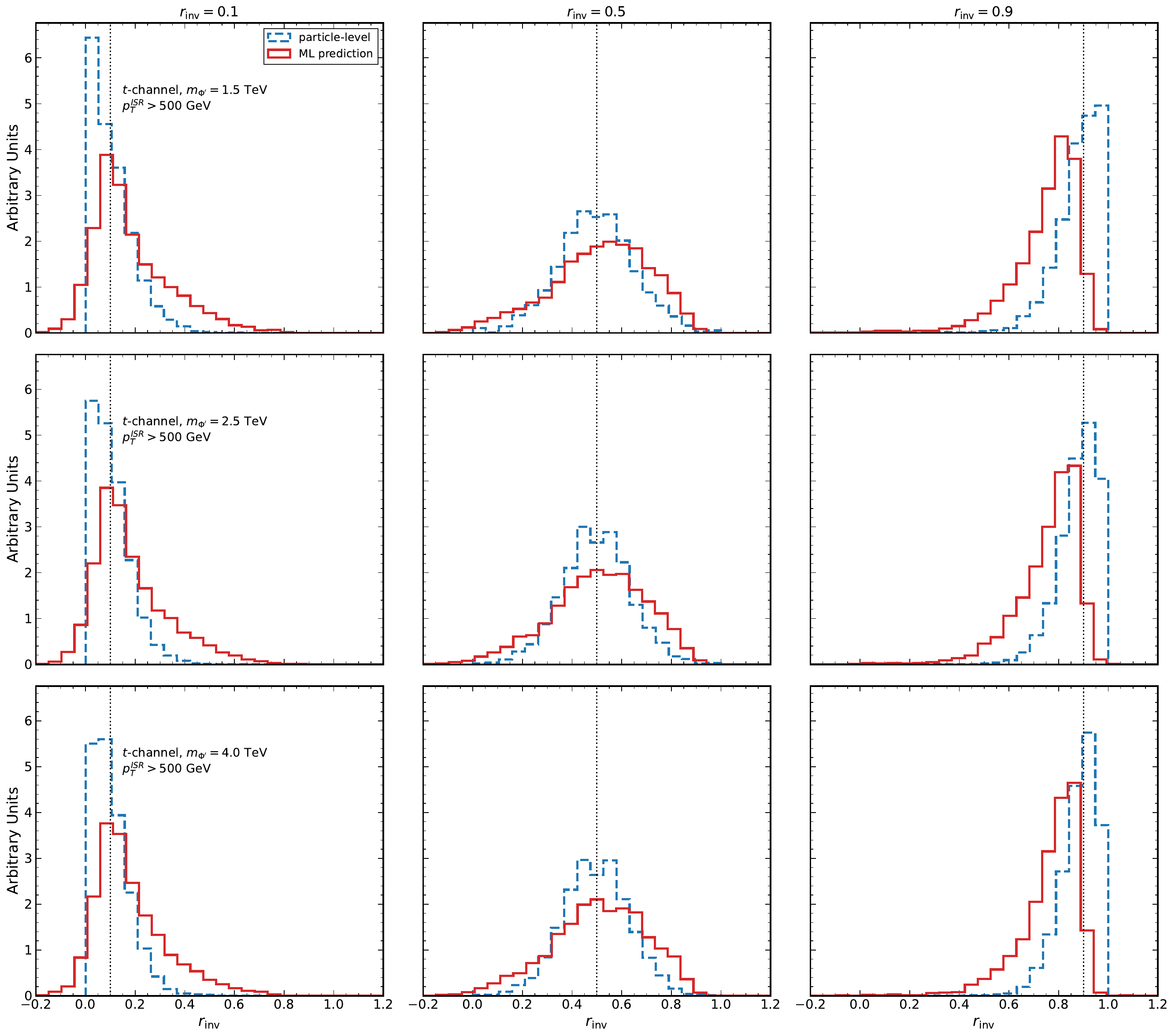}
    \caption{Distributions of \rinvML\ for $t$-channel samples with $m_{\phi}=1500$~\GeV\ (upper row) , $2500$\GeV(middle row) and  $4000$~\GeV\ (lower row), evaluated using the model trained on $s$-channel samples. From left to right, the panels correspond to $\rinv=0.1$, $0.5$, and $0.9$. The dashed histograms indicate the particle-level \rinv\ distributions.}
    \label{fig:tchannel}
\end{figure}

Although, the $t$-channel involves a scalar mediator $\Phi$ that couples an SM quark to a dark quark, very different to the $s$-channel production, the event-topology is very similar. Both production channels have a back-to-back SVJ system boosted by an ISR object. Given the same \rinv, the regression model should achieve similar performance for $t$-channel as well, even if the model is trained using $s$-channel samples only, which is confirmed in Fig.~\ref{fig:tchannel}.

Conventionally, the searches for $s$-channel and $t$-channel production are separated as the former relies on observables trying to reconstruct the resonance peak, which is not feasible for the other. The above finding is encouraging, since the regressed \rinv can be a potential powerful discriminant variable that is universally applicable to both production channels. We also want to point it out here that previous searches for $t$-channel SVJ places stringent selections to reduce the multijet background, resulting in a very different background composition, dominated by electroweak boson processes. \rinvML\ may provide sufficient separation power so that $t$-channel searches can loosen their selection criteria to have similar background components as the $t$-channel, which makes it feasible to consider both in a unified approach.  

\subsection{Analysis impact}
\label{sec:impact_ana}

\begin{figure}[th]
    \centering
    \includegraphics[width=0.85\linewidth]{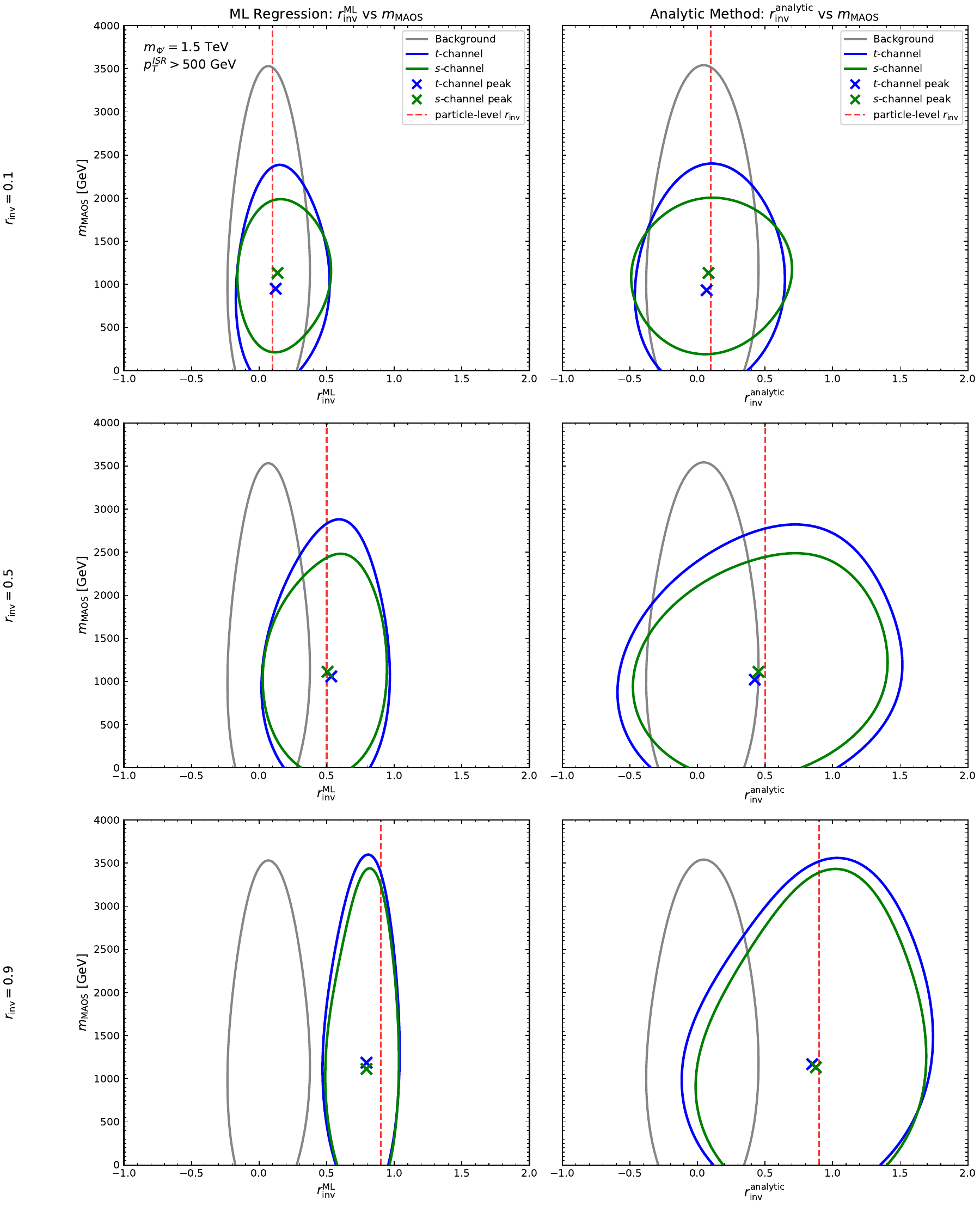}
    \caption{80\% contour regions in the two-dimensional $(\rinv,\maos)$ plane for the $t$-channel and $s$-channel samples at $m_{\phi}=m_{Z'}=1.5$~\TeV\ and $\PtISR>500$~\GeV, shown for $\rinv=0.1$ (top), $0.5$ (middle), and $0.9$ (bottom). The left column uses \rinvML, and the right column uses \rinvAna. In each panel, the grey contour denotes the background, the blue and green contours denote the $t$-channel and $s$-channel signals, respectively, and the markers indicate the corresponding density peaks. The red dashed line marks the particle-level \rinv\ value.}
    \label{fig:2d_scatter}
\end{figure}

Previous studies have shown that using \rinvAna\ with \maos, a mass variable built from the $M_{T2}$-assisted on-shell reconstruction of the invisible momenta, allows us to obtain much higher sensitivity~\cite{Cho:2008tj,Choi:2010dw,Liu:2024SVJplusX} for $s$-channel. The conclusion applies to \rinvML\ as well, given its improved \rinv\ precision. As shown in Fig.~\ref{fig:2d_scatter}, the signal events are more separated from background on the $\rinv - \maos$ plane when \rinvML\ is considered. It is also worth highlighting that both $s$-channel and $t$-channel exhibit similar clusters, while the peak is lower for $t$-channel in \maos.

The enhancement can be appreciated more via a quantified approach. We calculate the bin-wise $S/\sqrt{B}$ value in the two-dimensional plane, and Table~\ref{tab:s_over_sqrtb_compare} compares the maximum value obtained using analytic \rinv\ and \rinvML. It is striking that \rinvML brings significant gains, especially in the high \rinv region. 

\begin{table}[t]
\centering
\caption{Maximum $S/\sqrt{B}$ obtained in the two-dimensional $\rinv, \maos$ plane for the machine-learning-regressed and analytically reconstructed $\rinv$. The relative improvement is defined with respect to the analytic method.}
\label{tab:s_over_sqrtb_compare}
\begin{tabular}{c c c c}
\hline
particle-level \rinv & ML regression & Analytic reconstruction & Relative improvement \\
\hline
0.1 & 14.33 & 11.88 & 20.7\% \\
0.5 & 58.67 & 32.27 & 81.8\% \\
0.9 & 218.33 & 88.47 & 146.8\% \\
\hline
\end{tabular}
\end{table}

\clearpage
\section{Conclusion}
\label{sec:conclusion}
This work proposes a ML-based regression method to reconstruct \rinv, the key model parameter in SVJ searches. With the presence of an ISR-object, the model can achieve much better precision compared to the analytical formula developed in ref.~\cite{Liu:2024SVJplusX}. The training procedure is tailored to reduce the dependence on the signal model parameters, so the performance is robust against various conditions. The underlying physics captures the critical event-topology feature of the ISR + SVJ system, regardless of the production channels. It is found that the regression model trained using $s$-channel signal events also has similar performance when evaluating on $t$-channel processes. 

Even though we only included photon-ISR processes in this study, the results are still valid for other ISR-objects, as the overall event topology does not change significantly. In the case of jet-ISR, a practical challenge is how to distinguish the ISR-jet from the SVJs. It is shown that simple geometric selections can achieve acceptable performance~\cite{ATLAS:2024qqm,Liu:2024SVJplusX}. Naturally, if one has an excellent SVJ tagging algorithm, this problem is solved automatically~\cite{Luigi:SVJModels,Canelli:2021aps,Bernreuther:2020vhm}. It is interesting to further explore the connections between SVJ identification and \rinv\ reconstruction.   

This method has great application potential. Firstly, \rinvML\ can serve as a discriminant variable in SVJ searches, given the obvious separation power seen in Figure~\ref{fig:2d_scatter}. Furthermore, it motivates us to rethink about the search strategy as the performance is encouraging for both $s$-channel and $t$-channel. It is possible to loosen the selection criteria adopted in $t$-channel SVJ searches, broadening their coverage. Finally, this regression model is very suitable for reinterpretation~\cite{svj-contour}, since only high-level physics objects are included in the training. It will remain effective as long as the signal models under consideration are characterized by \rinv, no matter whether the SVJs are enriched with leptons~\cite{Cazzaniga:2022hxl,lep-svj,tau-svj}, heavy-flavour quarks~\cite{svj-b} or long-lived particles~\cite{svj-emj}. Finally, we want to point out that our method is not tied with how \rinv is realized in the generators, as it is calculated using the visible and invisible particles inside the jets directly.   

\section*{{Acknowledgments}}
The preliminary result was presented at the LHC Dark Showers Workshop in 2025, where we received several valuable comments. We thank the workshop organizers and participants. In particular, we thank Jackson Burzynski for suggesting $t$-channel studies. B.X. Liu and Y. Li are supported by Shenzhen Campus of the Sun
Yat-sen University under project 74140-12240013, and by the Young Scientists Fund (C Class) of the National Natural Science Foundation of China (Grant No.12405122). B.X. Liu is also supported by the Natural Science Foundation of Guangdong Province (Grant No.2026A1515012178). J.B. Wang, J.Q. Xie, K.R. Xu and R.H. Ye are supported by College Students' Innovative Entrepreneurial Training Plan Program, and they developed the training/validation pipeline. Z.H. Huang selected this topic for their undergraduate thesis, and developed the sample production workflow. 
\appendix
\section{Model and Training Details}
\label{sec:app}
This appendix provides additional technical details of the regression model used to estimate the event-level invisible fraction.  The model takes reconstruction-level observables as inputs, including the ISR photon, the two leading jets, the missing transverse momentum, and derived angular and kinematic variables. The input variables are preprocessed using a physics-motivated normalization procedure before training. 

The leading-jet transverse momentum is then used as the event-wise normalization scale. The photon transverse momentum, subleading-jet transverse momentum, missing transverse momentum, leading-jet mass, and subleading-jet mass are divided by the leading-jet transverse momentum. The leading-jet transverse momentum itself is used only as the normalization scale and is not retained as an independent input feature. After this physical normalization, the normalized transverse-momentum-like, mass, and missing-momentum variables are standardized using the mean and standard deviation of the training sample.  The azimuthal-angle and pseudorapidity variables are kept in their original forms.

The regression network is a fully connected feed-forward neural network. Each hidden layer is followed by batch normalization and a ReLU activation. The output layer consists of a single neuron corresponding to the predicted value of \rinv. The model is trained by minimizing the mean-squared error between the predicted \rinv and the particle-level target value defined in Eq.~\eqref{eq2}. The training uses the Adam optimizer, weight decay, a validation-based learning-rate scheduler, and gradient clipping. 

The network architecture and training hyperparameters are selected using an Optuna-based hyperparameter optimization~\cite{akiba2019optuna}. The validation loss is used as the optimization objective, so the different network depths used for the $\PtISR>150$\GeV and $\PtISR>500$\GeV selections are the result of validation-based hyperparameter selection rather than a manual architectural
choice. The final hyperparameters used in the analysis are summarized in Table~\ref{tab:ml_hyperparameters}.

\begin{table}[ht]
\centering
\begin{threeparttable}
\caption{Final hyperparameters selected for the \rinv regression models.}
\label{tab:ml_hyperparameters}
\begin{tabular}{lcc}
\hline
Hyperparameter 
& \(\PtISR>150\)\GeV 
& \(\PtISR>500\)\GeV \\
\hline
Hidden layers 
& \([256,128,64,32,16]\) 
& \([128,64,32,16]\) \\

Learning rate 
& \(10^{-3}\) 
& \(10^{-5}\) \\

Batch size 
& \(128\) 
& \(64\) \\

Maximum epochs 
& \(200\) 
& \(200\) \\


Weight decay 
& \(10^{-5}\) 
& \(10^{-5}\) \\
\hline
\end{tabular}
\end{threeparttable}
\end{table}

The code used for the Delphes-to-CSV conversion, \rinv prediction, and plotting workflow is publicly available at \url{https://github.com/YinLi28/svj_rinv_inference_demo}. The repository contains a lightweight inference workflow, including data preprocessing utilities, trained model loading, \rinv prediction, and example plotting scripts. Starting from Delphes-level inputs, the workflow converts the selected objects into the required tabular format, applies the same preprocessing as in the analysis, evaluates the trained regression model, and produces the corresponding \rinv distributions.
\clearpage

\bibliographystyle{JHEP}
\bibliography{main}

\end{document}